\documentstyle[nato,epsfig,graphicx,amsmath,amsfonts]{crckapb} 

\newcommand{\erfc}{{\mbox{erfc}}}
\newcommand{\infd}{{\mbox{d}}}

\newcommand{\ZZ}{{\Bbb{Z}}}

\newcommand{\VECk}{{\mathbf{k}}}

\newcommand{\VECm}{{\mathbf{m}}}

\newcommand{\VECr}{{\mathbf{r}}}

\begin{opening}

\title{Computer Simulations of charged systems}
\subtitle{}
\author{C. Holm, K. Kremer}

\institute{Max-Planck-Institut f\"ur Polymerforschung\\ 55128 Mainz, Germany}
\end{opening}

\begin{document}    

\begin{abstract}
  In this brief contribution to the Proceedings of the NATO-ASI on
  ``Electrostatic Effects in Soft Matter and Biophysics''\cite{les_houches},
  which took place in Les Houches from Oct. 1-13, 2000, we summarize in short
  aspects of the simulations methods to study charged systems. After
  describing some basics of Monte Carlo and Molecular dynamics techniques, we
  describe a few methods to compute long range interactions in periodic
  systems. After a brief detour to mean-field models, we describe our results
  obtained for flexible polyelectrolytes in good and bad solvents. We follow
  with a description of the inhomogeneity of the counterion distribution
  around finite chains, and continue then with infinitely long, rodlike
  systems. The last part is devoted to the phenomenon of overcharging for
  colloidal particles and its explanation in terms of simple electrostatic
  arguments.

\end{abstract}
\maketitle

\section{Introduction}\label{sec:introduction}

Polyelectrolytes\index{Polyelectrolytes} represent a broad and interesting
class of materials \cite{pe_review} that enjoy an increasing attention in the
scientific community.  For example, in technical applications polyelectrolytes
are wildly used as viscosity modifiers, precipitation agents, superabsorbers,
or leak protectors.  In biochemistry and molecular biology they are of
interest because virtually all proteins, as well as DNA, are polyelectrolytes.

In contrast to the theory of neutral polymer systems, which is well developed,
the theory of polyelectrolytes faces several difficulties.  Simple scaling
theories, which have been proven so successfully in neutral polymer theory,
have to deal with additional length scales set by the long range \index{Long
  range} Coulomb interaction\cite{joanny_lh}.  Furthermore there is a delicate
interplay between the electrostatic interaction of the distribution of the
counterions\index{Counterions} and the conformational degrees of freedom,
which in turn are governed by a host of short range interactions, which
renders the problem difficult.  There are only two limiting cases that are
easy to solve. These are the case of high salt excess, effectively screening
out the electrostatic interaction (which in turn allows one to treat it as a
perturbation), or the case of an overwhelming dominance of the Coulomb force,
which results in a strongly elongated chain.  Unfortunately it is often just
the intermediate case, which proves to be the most interesting regime in terms
of application, experiment and theory.

Computational simulations provide some unique ways to elucidate the properties
of charged systems. We first give a more general introduction to the relevant
simulation methods, and focus then on some recently obtained results.

\section{Simulations techniques}\label{sec:simulation}
\subsection{Some Basics on Simulations}
There exist a number of nice reviews and books
\cite{Binder:96.1,Kremer:033,Allen:89.1}
which deal extensively with various aspects of computer simulations of
complex, not necessarily charged, systems. Thus, the present introduction can
also be viewed as a guide to the literature.

There are two basic concepts, which are used in computer simulations
of complex systems.  The conceptionally most direct approach is the
molecular dynamics \index{Molecular dynamic}(MD) method. One
numerically solves Newton's equation of motion for a collection of
particles, which interact via a suitable interaction potential $U
(\vec{r}_{i})$, where $\vec {r}_{i}$ are the positions of the
particles.  Through the equation of motion a natural time scale is
built in, though, this might not be the physically realistic time
scale (e. g. if the solvent is replaced by a dielectric background).
Running such a simulation samples phase space for the considered
system deterministically. Though this sounds very simple, there are
many technical and conceptual complications, which one encounters on
the way. The second approach, the Monte Carlo\index{Monte Carlo} (MC)
method, samples phase space stochastically.  Monte Carlo is
intrinsically stable but has no natural time scale built in. This can
be reinterpreted, however, by an adjustment of suitable ''time
amplitudes''.  The MD and MC approaches are the basic simulation
methods for exploring the statistical properties of complex fluids. At
present, many applications employ variants thereof, or even hybrid
methods, where combinations of both are used. Before going into detail
we ask when is either kind of model appropriate?

At first sight it is tempting to perform a computer simulation of a
polyelectrolyte solution where all details of the chemical structure
of the monomers are included. For instance, the chain diffusion
constant D could be measured by monitoring the mean square
displacement of the monomers of the chains. This, however, is tempting
only at the very first glance.  Even for the fastest computers one
would need an exceeding amount of computer time.  As for all
disordered, complex, macromolecular materials, (charged) polymers are
characterized by a hierarchy of different length and time scales, and
these length and especially the time scales span an extremely wide
range \cite{Baschnagel:97.1}.  On the atomistic level the properties
are dominated by the local oscillations of bond angles and
lengths.\footnote{For reactions or to study exited states, the
electronic structure is treated explicitely. Such methods
(Carr-Parrinello simulations, quantum chemistry etc.) are beyond the
scope of the present paper.}  The typical time constant of about
$10^{-13}$ sec results in a simulation time step of $10^{-15}$ sec.
On the semi-macroscopic level the behavior is dominated by the overall
relaxations of conformation of the objects or even larger units
(domains etc).  The times, depending on chain length, temperature and
density, can easily reach seconds.  To cover that many decades in time
within a conventional computer simulation is certainly impossible at
present. On the other hand, it is important to relate the chemical
structure of a system to its properties.

%
\subsection{Molecular Dynamics}
%
MD simulations date back to the early fifties. For a rather complete
overview about simulations in condensed matter we refer to
\cite{Binder:96.1}. Consider a cubic box of Volume $V = L^{3}$,
containing $N$ identical particles. In order to avoid surface effects
and (as much as possible) finite size effects, one typically uses
periodic boundary conditions. The particle number density is given by
$\rho = N / L^{3}$. The first simulations employed hard spheres of
radius $R_{o}$, leading to a volume fraction $\rho _{v} = 4/3 \pi
R^{3}_{o} \rho$.  Though still used extensively for some studies on
the glass transition of colloidal systems we focus here on soft
potentials.

A key thermodynamic quantity, the temperature, is imposed via the
equipartition theorem $\frac {m}{2} \langle {\vec{\dot r}}^{2}_{i}
\rangle = \frac {3}{2} kT$, $m$ being the particle masses. Note that
in hard sphere systems temperature only defines a time scale, but is
otherwise irrelevant. One can find many different soft potentials in
the literature.  However, most widely used is the Lennard-Jones
potential $ U^{LJ} ({r}_{ij})$, derived originally for interactions of
noble gases (Ar, Kr \dots), ${r}_{ij}$ being the distance between
particle $i$ and $j$. In its simplest form for two identical particles
it reads
\begin{eqnarray}
U^{LJ} ( {r}_{ij}) = 4 \epsilon  \left[ (\frac
{\sigma}{r_{ij}})^{12} - (\frac {\sigma}{r_{ij}})^{6} \right]
\end{eqnarray}
Usually, a cutoff $r_{c}$ is introduced for the range of the
interaction. This typically varies between 2.5 $\sigma$ (classical LJ
interaction with an attractive well of depth $\epsilon$, used for the
poor solvent\index{Poor solvent} chains in
Sec.~\ref{subsec:poorsolvent}) and 2$^{1/6} \sigma$ (the potential is
cut off at the minimum, leading to a pure repulsive interaction, as is
typically done for good solvent chains).  For chain molecules a
bonding interaction for $r < R_0$
 \begin{eqnarray}
 U_{\rm FENE}(r) = - \frac {1}{2}  k R_{0}^{2} \ln (1 - \frac {r^{2}} {R_{0}^{2}})
 \end{eqnarray}   
 is added, which keeps the bond length below a maximum of $R_{0}$.  The spring
 constant $k$ varies between 5 and 30 $\epsilon / \sigma^{2}$. Electrostatics
 is included via the Coulomb interaction
\begin{eqnarray}
U_{c} (r_{ij}) = \frac {l_{B} k_{B} T} {r_{ij}} q_{i} q_{j}
\end{eqnarray}
Here $l_{B} := \frac{e_0^2}{4 \pi \varepsilon_0 \varepsilon_r k_B T}$,
where $e_0$ is the elementary unit charge, $k_B$ is the Boltzmann constant,
$T$ denotes temperature, $\varepsilon_0$ and $\varepsilon_r$ are the vacuum
and relative dielectric permeability of the solvent, respectively, and $q_{i,
  j}$ are charges measured in units of $e_0$.  Two monovalent charges
separated by the Bjerrum length $l_{B}$ have an interaction energy equal
to $k_B T$. The Bjerrum length thus is a measure of the interaction strength.
It is equal to 7.14 {\AA} ~for water at room temperature. The computational
aspects of the long range potential will be discussed shortly in
Sec.\ref{subsec:longrange}.

The unit of energy is $\epsilon$, of length $\sigma$ and of mass $m$.
This defines the ''LJ-units'' for temperature $[T] = \epsilon/(k_B
T)$, time $[t] = \sqrt {\sigma^{2} m/\epsilon}$ and number density
$[\rho] = \sigma^{-3}$. In most practical programs $\sigma$, $m$,
$\epsilon$ are used as the basic units and set to one. The straight
forward simulation technique is to integrate Newton's equations of
motion for the particles:
\begin{eqnarray}
m_{i} \ddot {\vec {r}}_{i} = - \vec \nabla \sum_{j, j
\not= i} U ({r}_{ij})
\end{eqnarray}
Since energy in such a simulation is conserved we have a
microcanonical ensemble. Presently other thermodynamic ensembles are
commonly used for practical applications (NPT: isobaric-isothermal,
NVT: isothermal (canonic) \dots).  Because we often employ a
stochastic MD method, known as the Langevin thermostat\cite{GrKr86}, we
will briefly describe its main ingredients.  Instead of
integrating Newton's equations of motion, one solves a set of Langevin
equations
\begin{eqnarray}
  m_i \ddot{\vec r}_i \; = \; -\vec \nabla \sum_{j, j\not= i} U ({r}_{ij}) 
- \Gamma \dot {\vec r}_i +  {\vec \xi}_i(t)
  \label{eq:Langevin_equation}
\end{eqnarray}
with ${\vec \xi}_i(t)$ being a $\delta$-correlated Gaussian noise source with
its first and second moments given by
\begin{eqnarray}
  \langle {\vec \xi}_i(t) \rangle = 0 
  \mbox{~~~~~~and~~~~~~} 
  \langle {\vec \xi}_i(t)\cdot {\vec \xi}_j(t') \rangle = 
  6 \, k_B T \, \Gamma \delta_{ij} \delta(t-t').
  \label{eq:while_noise_moments}
\end{eqnarray}
The friction term $-\Gamma\dot{\vec r}_i$ and the noise ${\vec
\xi}_i(t)$ is thought of as imitating the presence of a surrounding
viscous medium responsible for a drag force and random collisions,
respectively.  The second moment of ${\vec \xi}_i(t)$ is adjusted via
an Einstein relation in order to reach the canonical state in the
limit $t\rightarrow\infty$.  The dynamics generated by the Langevin
equation can alternatively be written as a general Fokker-Planck
process. This permits a transparent proof of two important facts: {\em
(i)\/} the stationary state of the process is the Boltzmann
distribution and {\em (ii)\/} the system will evolve and converge to
the Boltzmann distribution \cite{Ris89}.  Since for small times the
stochastic part is more important than the deterministic one
($\sqrt{t}\gg t$ for small $t$) it is actually not necessary to use
Gaussian random variables in the simulation \cite{DuPa91}. It suffices
to use equidistributed random variables with first and second moment
being identical to the Gaussian deviate.

A simple but very efficient and stable integration scheme is the
Verlet algorithm (more complicated methods, which do not have time
inversion symmetry, do not, in general, perform significantly
better). With a simulation time step $\delta t$ , where $\delta t << 2
\pi / \omega_{max}$ and $\omega_{max}$ is the typical highest
frequency of the system (for crystals the Einstein frequency), we have
in one dimension
\begin{eqnarray}
  r_{i} (t + \delta t) &=& r_{i} (t) + \delta t {v}_{i} (t) +
\frac {\delta t^{2}} {2} a_{i} (t) + \frac {\delta t^{3}}
{6} \dot {a_{i}} (t) + {\cal O}(\delta t^4) \nonumber \\
 {r}_{i} (t - \delta t)  &=&  {r}_{i} (t) - \delta t v_{i}
(t) + \frac {\delta t^{2}} {2} a_{i} (t) - \frac {\delta
t^{3}} {6} \dot a_{i} (t) + {\cal O}(\delta t^4) ,\nonumber 
\end{eqnarray}
where $v_i (t) = \dot r_i(t)$ and $a_i(t) = \dot v_i(t)$. An
addition of the two lines yields
\begin{eqnarray}
r_{i} (t + \delta t) = 2 r_{i} (t) - r_{i} (t - \delta t) + \delta
t^{2} a_{i} (t) + {\cal O} (\delta t^{4})
\end{eqnarray}
Therefore the position calculations have an algorithmic error of ${\cal O}
(\delta t^{4})$. Subtraction of the lines yields
\begin{eqnarray}
v_{i} (t)  = \frac {1} {2 \delta t} [r_{i} (t +
\delta  t) - r_{i} (t - \delta t) ] + {\cal O}(\delta t^{3})
\end{eqnarray}
leading to errors of $ {\cal O} (\delta t^{3})$. There are many
variants of this basic integrating scheme used throughout the
literature \cite{Allen:89.1}.  One can follow the realistic time
evolution of a system, as long as the forces/potentials are realistic
for the modeled system and as long as classical mechanics is
sufficient. In a purely deterministic simulation the accumulation of
small errors can cause significant deviations from the real
trajectory.  If the system is ergodic, which requires mixing of normal
modes (recall the well-known Fermi-Pasta-Ulam problem, where one asks
how anharmonic a potential has to be in order to equilibrate a one
dimensional chain of particles\cite{fermi55}) one can determine
ensemble averages from time averages
\begin{eqnarray}
<A>  = \frac {1}{M} \sum^{M}_{i = 1} A (t_{i})
\end{eqnarray}
of any physical quantity $A$ of interest. This describes the most
elementary Ansatz for a microcanonical simulation
\cite{Allen:89.1}. Here all extensive thermodynamic variables of the
system, namely $N$, $V$, $E$ are kept constant.  Sometimes this is
also called $NVE$ ensemble. As mentioned before, most applications
employ other ensembles such as the canonic ($NVT$), the
isobaric-isothermal ($NPT$) or even the grand canonical $(\mu, P, T)$
ensemble, $\mu$ being the chemical potential. As a general rule, in
cases such as two phase coexistence or calculations of transport
properties, one should choose an ensemble with many intensive
variables kept constant as possible. For charged systems, however, it
is rather difficult to perform efficient simulations in the ($N$, $P$,
$T$) or ($\mu$, $P$, $T$) ensembles.  Therefore the most common
ensemble is the NVT since it is easy to use the deterministic
equations with additional stochastic terms to constrain the
temperature (Eq. \ref{eq:Langevin_equation}).

%
%
\subsection{Monte Carlo Method}
%
The classical Monte Carlo approach goes to the other extreme, namely
to purely stochastic sampling.  Starting from a particular
configuration, randomly a particle (or a number of particles) is
selected and displaced by a random jump.  For hard sphere systems the
move is accepted if the new configuration complies with the excluded
volume; if not the old configuration is retained.  The approach is
also called simple sampling. This cycle is repeated over and
over. Once every particle on average has a chance to move, one Monte
Carlo step is completed. This is the most basic Monte Carlo simulation
(see e.g.\cite{Binder:96.1,Binder:79.1}). Since there is no energy
involved, it trivially fulfills detailed balance
\begin{eqnarray}
W (\lbrace x \rbrace \rightarrow \lbrace y \rbrace)  P_{eq} (
\lbrace x \rbrace) = W (\lbrace y \rbrace \rightarrow \lbrace x
\rbrace) P_{eq} ( \lbrace y \rbrace )
\end{eqnarray}
where $ W ( \lbrace x \rbrace \rightarrow \lbrace y \rbrace )$ is the
probability to jump from state $ \lbrace x \rbrace$ to state $\lbrace y
\rbrace$ and $P_{eq} (\lbrace x \rbrace)$ the equilibrium probability of state
$ \lbrace x \rbrace$. All states have exactly the same probability. Detailed
balance is a sufficient condition for a MC simulation to relax into thermal
equilibrium, though this may take a very long time. Special cases of
algorithms without detailed balance will not be discussed here.

It is useful to compare the basic aspects of MC simulation to the
examples discussed above for molecular dynamics simulations. The
Hamiltonian depends for simplicity only on the positions of all
particles $\{r_{i} \}$ and is denoted by $H ( \{r_{i}\}$.  The
expectation value of any observable A is given by
\begin{eqnarray}
<A> = \sum_{\lbrace r_{i} \rbrace} A( \lbrace r_{i} \rbrace) P_{eq} (
\lbrace r_{i} \rbrace)
\end{eqnarray}
with
\begin{eqnarray}
  P_{eq} ( \lbrace r_{i} \rbrace) &=& \exp (- H (\lbrace r_{i} \rbrace)
/ k_{B}T)/Z \nonumber \\ Z &=&  \sum_{\lbrace r_{i} \rbrace} \exp (-H
/ k_{B} T)
\end{eqnarray}
An exact way would be to sample all possible states, which in all but
the most trivial cases is impossible. Thus we sample phase space
stochastically.  Taking a particle at random, calculating its energy,
one moves it and calculates the new energy.  With $P (\lbrace x
\rbrace )$ being the Boltzmann-probability of the original state and $
P (\lbrace y \rbrace)$ being that of the new state, detailed balance
is obeyed if
\begin{eqnarray}
\frac {W (\lbrace x \rbrace \rightarrow \lbrace y \rbrace)} {W
(\lbrace y
  \rbrace \rightarrow \lbrace y \rbrace)} = \exp \lbrace - (H (\lbrace x
\rbrace) - H (\lbrace y \rbrace ) /k_{B}T) \rbrace
\end{eqnarray}
Under this condition the algorithm is ergodic and the system relaxes into
equilibrium. The Metropolis method is the most frequently used prescription
one to accept or reject a move:
\begin{eqnarray}
W (\lbrace x \rbrace \rightarrow \lbrace y \rbrace) = \Gamma \Bigg
\lbrace
\begin{array}{ccc}
\exp (H (\lbrace x \rbrace) - H (\lbrace y \rbrace)  / k_{B}T)
&,&\triangle H > 0\\ 1&,&\triangle H < 0.
\end{array}
\end{eqnarray}
Since only the ratio of the W's is relevant, $\Gamma$ is an arbitrary
constant between zero and one, usually $\Gamma$ = 1. A random number
$x$, equally distributed between 0 and 1, is used to decide upon the
acceptance of a move.  If $x < W (\lbrace x \rbrace \rightarrow
\lbrace y \rbrace )$ the move is accepted, otherwise rejected. (For
$\Gamma$ = 1 any move, which lowers the energy is accepted.) This is
the basic MC procedure used for sampling phase space in statistical
physics.

In many cases, however, one also would like to gain information on the
dynamics of a system or even better, of a model system. How can a MC
simulation, with no intrinsic time scale, be used to obtain
information on the dynamics? In the method described above the system
evolves from one state to another by a {\bf local} move. Through these
local stochastic moves the configurations of particles change with
''time''.  This is a dynamic MC method based on a Markov process,
where subsequent configurations $ \lbrace x \rbrace \rightarrow
\lbrace x^{i} \rbrace \rightarrow \lbrace x^{ii} \rbrace \rightarrow
\dots$ are generated with a transition probability $ W (\lbrace
x^{i}\rbrace \rightarrow \lbrace x ^{ii} \rbrace)$. To a large extent
the choice of the move is arbitrary, as long as one can interpret it
as a local elementary unit of motion.  The prefactor $\Gamma$ can
actually be interpreted as an attempt rate $\Gamma = \tau_{o}^{-1}$
for the moves and introduces a timescale. This ''changes'' the purely
statistical transition probability into a transition probability per
unit time\cite{Kremer:033,Binder:79.1}. To compare simulated
(overdamped) dynamics with an experiment, it essentially requires
determination of $\tau_{o}$ (e. g. diffusion constants). It is
obvious, that this simulation does {\bf not} include any hydrodynamic
effects since there is no momentum involved. There are very
interesting, more advanced methods like DPD (dissipative particle
dynamics) and Lattice-Boltzmann methods, currently under development
in order to include this efficiently \cite{EsWa95}). Using the
interpretation of a MC step as a time step, ensemble averages can be
written as time averages:
\begin{eqnarray}
<A> = \frac {1}{M - M_{o}} \sum^{M}_{i=M_{o} + 1} A (\lbrace x^{i}
\rbrace) \cong \frac {1} {t - t_{o}} \int^{t}_{t_{o}} dt' A (t') ~.
\end{eqnarray}
We view one attempted move per system particle as one
time-step.

The first configurations in a simulation are usually not yet
equilibrium configurations. One first has to ''relax'' the system into
equilibrium, meaning the data for the first $M_{o}$ steps are
omitted. In this interpretation the dynamic Monte Carlo procedure is
nothing but a numerical realization of a Markov process described by a
Markovian master equation
\begin{eqnarray}
\frac {d}{dt} P (\lbrace x \rbrace, t) &=& - \sum_{\lbrace x^{i}
\rbrace} W (\lbrace x \rbrace \rightarrow \lbrace x^{i} \rbrace) P
(\lbrace x \rbrace, t)\nonumber\\ &&+ \sum_{\lbrace x^{i} \rbrace} W (\lbrace
x^{i} \rbrace \rightarrow \lbrace x \rbrace) P (\lbrace x^{i}
\rbrace, t)
\end{eqnarray}
with $P (\lbrace x \rbrace, t)$ the time dependent probability of state
$\lbrace x \rbrace$. The condition of detailed balance is sufficient that
$P_{eq} (\lbrace x \rbrace)$ is the steady-state solution of the master
equation. If all states are mutually accessible $P (\lbrace x \rbrace, t)$
must relax towards $P_{eq} (\lbrace x \rbrace )$ as $ t \rightarrow \infty$
irrespective of the starting state. Note however that the choice of a ''good''
starting state can save enormous amounts of CPU time.

So far, the two extreme cases for classical, particle based computer
simulations were discussed, microcanonical MD and canonical MC.  There
are many approaches ''in between'' which are used depending on the
problem under consideration. The techniques range from pure MD, where
Newton's equations of motion are solved ($\ddot x = - \nabla U)$, MD
coupled to a heat bath and added friction (''Langevin MD'', ''Noisy
MD''), ($\ddot x = - \nabla U - \zeta \dot x + f (t))$, $\zeta$
friction, f (t) random force), Brownian Dynamics (BD) ($\dot x = -
\nabla U$ + random displacement), force biased MC (attempted moves are
selected from the very beginning according to local forces), to plain
MC as described above.

For the application to polymers one should keep in mind, that the
conformational entropy of the chains add additional complications, which make
proper equilibration especially difficult or time consuming. Thus, wherever
possible, methods should be used, which are faster than the slow intrinsic
dynamics of the chains \cite{Kremer:033}.

\subsection{Methods for long range interactions}\label{subsec:longrange}
One of the biggest problems for the simulations of charged systems is
the long range nature of the Coulomb interactions. In principle, each
charge interacts which all others, leading to a computational effort
of ${\cal O}(N^2)$ already within the central simulation box.  For
many physical investigations one wants to simulate bulk properties and
therefore introduces periodic boundary conditions to avoid boundary
effects. The standard method to compute the merely conditionally
convergent Coulomb sum
\begin{eqnarray}
  E = \frac{1}{2} \sum_{\vec n}^{~~~~\prime} \sum_{ij}
\frac{q_i q_j}{|\vec r_{ij} + \vec n L|},\label{eq:coulomb}
\end{eqnarray}
where the prime denotes that for $\vec n = \vec 0$ the term $i=j$ has
to be omitted, is the traditional Ewald summation \cite{ewald21a}. The
basic idea is to split the original sum via a simple transformation
into two exponentially convergent parts, where the first one,
$\phi_r$, is short ranged and evaluated in real space, the other one,
$\phi_k$, is long ranged and can be analytically Fourier transformed
and evaluated in Fourier space:
\begin{eqnarray}
  \frac{1}{r} = \frac{1-f(r)}{r} + \frac{f(r)}{r}\simeq \phi_r (r_c,
  \alpha) + \phi_k(k_c,\alpha )
\end{eqnarray}
Traditionally, one uses for $f$ the error function ${\rm erf}(\alpha
r):=2\pi^{1/2}\int_0^{\alpha r} \exp{-t^2}\break {\rm d}t$, though
other choices are possible and sometimes more
advantageous\cite{heyes80a,berendsen93a,huenenberger00a}.  For any
choice of the Ewald parameter $\alpha$ and no truncation in the sums
the formula yields the exact result.  In practice one wants to cut off
the infinite sum at some finite values $r_c$ and $k_c$ to obtain $E$
to a user controlled accuracy, which is possible by using error
estimates \cite{kolafa92a}.  The aformentionend procedure results in
the well known Ewald formula for the energy of the box
\begin{eqnarray}\label{EwaldAnteile}
E = E^{(r)} + E^{(k)} + E^{(s)} + E^{(d)},
\end{eqnarray}
where the contributions from left to right are the real space, 
Fourier space, self , and dipole-correction energy terms. These are
given by
\begin{eqnarray}
E^{(r)} & = & \frac{1}{2} \sum_{i,j} \sum_{\VECm\in\ZZ^{3}}^{\prime}
q_{i}q_{j} \frac{\erfc(\alpha|\VECr_{ij}+\VECm L|)}
{|\VECr_{ij}+\VECm L|} \label{Realraumanteil} \\
E^{(k)} & = & \frac{1}{2}\frac{1}{L^{3}} \sum_{\VECk\ne 0}
\frac{4\pi}{k^{2}} e^{-k^{2}/4\alpha^{2}}
|\tilde{\rho}(\VECk)|^{2} \label{Impulsraumanteil} \\
E^{(s)} & = & -\frac{\alpha}{\sqrt{\pi}}\sum_{i}q_{i}^{2}
\label{Selbstenergie} \\
E^{(d)} & = &
\frac{2\pi}{(1+2\epsilon')L^{3}}\left(\sum_{i}q_{i}\VECr_{i}\right)^{2},
\label{Dipolkorrektur}
\end{eqnarray}
and the Fourier transformed charge density $\tilde{\rho}(\VECk)$ is defined as
\begin{eqnarray}\label{FLadungsdichte}
\tilde{\rho}(\VECk) = 
\int_{V_{b}}\infd^{3}r \; \rho(\VECr)e^{-i\;\VECk\cdot\VECr} =
\sum_{j=1}^{N}q_{j}\;e^{-i\,\VECk\cdot\VECr_{j}}.
\end{eqnarray}
The dipole term depends not on $\alpha$, hence is independent of the
splitting function. It reflects the way Eq. (\ref{eq:coulomb}) is
summed up, here in a spherical way towards
infinity\cite{deleeuw80a}. It also includes a correction for the
dielectric constant $\epsilon ^\prime$ outside the summed up sphere
volume.  For metallic boundary conditions, $\epsilon ^\prime =
\infty$, and the dipole term vanishes. In principle the thermodynamic
properties should be independent of the choice of boundary
conditions\cite{deleeuw80b}. The Ewald sum has complexity ${\cal
O}(N^{3/2})$ in its optimal implementation \cite{perram88a}, and
therefore is not suitable for the study of large systems ($N > {\cal
O} (1000)$).  Implementing a fast Fourier transformation (FFT) for the
Fourier part results in the so-called particle-mesh-Ewald
formulations, which improve the efficiency to ${\cal O} (N \log N)$
\cite{hockney88a,darden93a,essmann95a,petersen95a}, and which can also
be efficiently be parallelized\cite{pollock96a,limbach01c}. The most
versatile and accurate method of all mesh-methods is the oldest P3M
algorithm\cite{hockney88a,deserno98a}, for which also precise error
estimates exist\cite{deserno98b}.  Another way of computing
Eq.(\ref{eq:coulomb}) is via a convergence factor
\begin{eqnarray}
  \label{eq:convergence}
  E = \lim_{\beta \rightarrow 0 } \frac{1}{2} \sum_{\vec n}^{~~~~\prime} \sum_{ij}
\frac{q_i q_j \exp{ (-\beta |\vec r_{ij} + \vec n L|) }}{|\vec r_{ij} + \vec n
  L|}.
\end{eqnarray}
This approach is used in the Lekner \cite{lekner91a} and Sperb
\cite{sperb98a} methods to efficiently sum up the 3D Coulomb sum.
Although the method in its original versions has ${\cal O} (N^2)$
complexity, Sperb et al. have developed a factorization approach which
yields an ${\cal O} (N \log N)$ algorithm \cite{sperb01a}.

Other advanced methods of ${\cal O} (N \log N)$ are tree
algorithms\cite{barnes86a}, which are the first order approximation of
even better, so-called fast multipole methods
\cite{greengard87a}. These can reach a linear complexity, but at the
expense of a heavy computational overhead which makes these methods
advantageous only for a very large number of charges ($N \approx
100\,000$) \cite{esselink94a}.

For thin polyelectrolytes films or membrane interactions one is also
interested in summations where only 2 dimensions are periodically
replicated and the third one is of finite thickness $h$ ($2D + h$
geometry). For this geometry Ewald based formulas are only slowly
convergent, have mostly ${\cal O}(N^2)$ scalings and no ``a priori''
error estimates exist \cite{widmann97a}. Recently Arnold
\cite{arnold01a,arnold01b} developed a method which is based on
convergent factors, whose errors are well controlled, and which uses a
factorization approach resulting in an ${\cal O}(N^{5/3})$ scaling
(MMM2D). In two dimensions the convergence factor based methods and
the Ewald sum methods yield exactly the same results, there is no
dipolar correction term needed\cite{arnold01a}. This is in contrast to
the 3D methods\cite{deleeuw80a}.  However, an even better scaling can
be achieved, if one returns to the 3D Ewald formula, and allows for a
large empty space between the unwanted replicas in the third dimension
\cite{yeh99a}. However, so far the method has been only checked on a
trial and error basis. We recently improved this situation by
computing an analytic error term which accounts for the contributions
of the unwanted replicas. By simply subtracting this term from the 3D
sum, which is a linear operation in $N$, one can in principle come as
close as desired to the real $2D +h$ sum, by allowing for just some
arbitrary small amount of empty space between the
layers\cite{arnold01c}. Using then again the P3M method we obtain an
$N\log N$ scaling with well controlled errors also for the $2D + h$
geometry, which up to now seems to be the optimal choice.

To simulate the structure of water (or other dipolar solvents), one
also needs to treat the dipolar interactions in a similar
fashion. Also here the Ewald method is applicable\cite{perram88a}, and
error estimates exist\cite{wang01a}.

\section{Mean-field models\index{Mean-field}: Debye-H\"uckel chains\index{Debye-H\"uckel chains}}\label{sec:mean-fieled}

\begin{figure}[tb]
  \begin{center}
    \leavevmode
    \epsfig{file=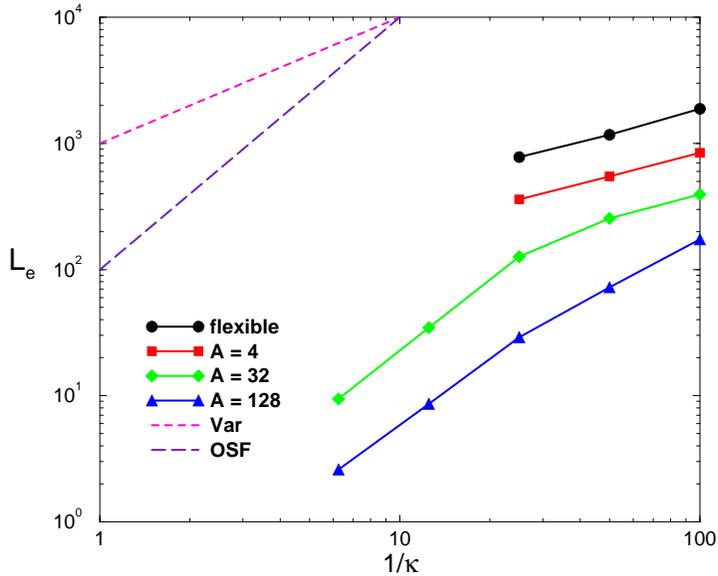, width=0.8\textwidth}
\caption{Electrostatic persistence length as a function of
$1/\kappa$ for several $A$ values.}
    \label{mickaf:fig2}
  \end{center}
\end{figure}

While we deal here mainly with systems where the charges are
explicitly taken into account, historically and even up to now many
studies consider the ions solely in a mean field approximation. In the
first step all non bonded charges are considered as a smeared
continuous charge density.  Such a situation is described by the
Poisson-Boltzmann\index{Poisson-Boltzmann} (PB) equation. While this
in most cases is not exactly solvable, many studies employ the
Debye-H\"uckel approximation, which is the solution of the linearized
PB
equation\cite{deserno_lh,kjellander_lh,moreira_lh,podgornik_lh}. The
resulting potential between charges is the screened Coulomb potential
\begin{eqnarray}
V_{DH} = \frac{l_{B}}{e_0}  k_{B} T \frac {\exp (- \kappa
r)} {r},
\end{eqnarray}
with$1 / \kappa$ being the Debye screening length. The polymer can now
easily be modeled as a random walk of $N$ monomers from which a
fraction $f$ is monovalently charged. If one in addition introduces
the stiffness along the backbone of the chain by a cosine-potential,
the total Hamiltonian reads
\begin{eqnarray}
\frac {H} {k_{B} T} = - A \sum^{N-1}_{i = 1}  (\vec{b}_{i}\cdot
\vec{b}_{i+1}) + \sum ^{N}_{i = 2} \sum^{i - 1}_{j = 1}
 \theta (q_{i} q_{j}) l_B \frac {\exp (- \kappa r_{ij})} {r_{ij} k_B T} .
\end{eqnarray}
The positive amplitude $A$ defines the strength of the angular
potential and the Heavyside $\theta$-function is one if the monomers
$i$ and $j$ are charged, and zero otherwise. The symbol $\vec b_{i}$
is the bond vector between monomer number $i$ and $i + 1$. For the
present simulations the bond length $|\vec b |$ and the Bjerrum length
$l_B$ are fixed to one $\sigma$, and thus set the basic length
scale. One bond mimics several neutral monomers as usual for coarse
grained simulation models. The longest chains considered contained up
to 2049 repeat units with a charge fraction of $f = \frac {1} {16}$.
Mapping this onto a PS-NaPSS copolymer for a more flexible case and
keeping in mind that for these polymers roughly three repeat units fit
into one Bjerrum length one arrives at a molecular weight of more than
600\,000 g/mol.  Thus, for these kind of questions computer
simulations are quite capable of covering the experimentally
interesting regime. This allows us to systematically vary not only the
chain length and the screening length but also the chain bending
stiffness. This is only of limited experimental relevance since very
long isolated chains in dilute solutions cannot be experimentally
analyzed so far. However one of the central questions in the theory of
polyelectrolytes is whether the characteristic electrostatic length is
a linear or quadratic function of the Debye length (i.e. proportional
to the square of charge density or to the charge density
itself). Analytic results mainly predict a $\kappa^{2}$ asymptotic
dependency of the persistence length\cite{joanny_lh}. However, an
estimate of the required the chain length to reach the asymptotic
regime for flexible weakly charged polyelectrolytes shows that the
chains are so long that all experimentally realizable concentrations
are in the semi-dilute regime. Thus, this is a typical question of
interest which is of no direct importance for experiments.
Nevertheless, it can add significantly to our understanding which in
turn can also influence the interpretation of experiments in the
semi-dilute regime.  Therefore, it is worthwhile to undertake some
efforts in computer simulation to investigate this question. There are a
number of attempts to do this, which so far do not lead to a clear-cut
answer. A typical result from such a simulation is given in
Fig.~\ref{mickaf:fig2}.

It shows the dependence of the electrostatic persistence length on the
screening length. The two dotted lines indicate the variational (Var)
and the Odijk-Skolnik-Fixman (OSF) results ( $L_{e} \propto
\kappa^{-1},$ or $\kappa^{-2}$ respectively). As one can see for the
regime covered there seems to be a continuous transition from a very
weak dependency on $\kappa$ towards the asymptotically expected
$\kappa^{-2}$ behavior for the semi-flexible polyelectrolyte. In the
limit of very stiff chains the OSF result becomes exact and the data
has to agree with that. Even though the simulated chains lie very well
within the experimentally relevant regime, there is, at least for
typical experimental flexibilities, no clear sign of the crossover
into the asymptotic OSF regime. This allows for two important
conclusions. First of all, the chains are certainly not long enough to
display asymptotic behavior. Secondly, typical experimental chains are
also not long enough to display the asymptotic behavior as predicted
by mean-field\index{Mean-field} theories. Unlike neutral polymers,
polyelectrolytes normally are not in an asymptotic limit where the
predicted scaling laws can be cleanly observed. Another consequence is
revealed in a closer analysis of the simulation data. For semiflexible
polyelectrolyte chains there is no unique persistence length anymore,
as all theoretical pictures assume. Over short distances the intrinsic
stiffness dominates and gives a clear signal in the bond direction
correlation function. Only over large distances along the backbone of
the chain does the electrostatic effect show up and introduces a
second characteristic length scale into the system.  This can lead to
the special situation that the electrostatic contribution to the
persistence length for stiff chains is smaller than for flexible
chains. The reason simply lies in the fact that for the stiff chains
the charges along the backbone are already much further apart because
of the intrinsic stiffness as opposed to flexible chains. These larger
distances then experience a dramatically weaker electrostatic
repulsion due to the screening of the electrostatic interaction
resulting in a weaker effect on the persistence length. For details we
refer to Ref. \cite{mickac3}.

Altogether, the main conclusion from these simulations of simplified
mean-field like models is on the one hand that polyelectrolytes have
to be extremely long to be in the asymptotic regime, and that one has
to be very careful in deriving general statements from either
simulations, which typically are not asymptotic just as experiments,
and analytic theory, which is usually in an asymptotic limit.

Another conclusion is that the concept of persistence length for
polyelectrolytes certainly is not well defined in terms of the
properties of the chains. As soon as intrinsic stiffness is included
there is no longer a unique length scale that describes the internal
structure of the chain. On the other hand this is the basis of all
classical models used in analytic theory.

\section{Solutions of flexible polyelectrolytes}\label{sec:pesolutions}

\subsection{Good solvent chains with explicit counterions}

The first investigation of totally flexible many chain polyelectrolyte
systems in good solvent with explicit monovalent counterions was
performed several years ago\cite{stevens95a}.  The simulations were
carried out mostly with systems of 8 or 16 chains with $N_{m}$ = 16,
32 and 64. Instead of the P3M algorithm a spherical approximation in a
truncated octahedral simulation box was used which, for values smaller
than $N_{total} \approx 500$ is faster than the PME method. More
details of the whole study can be found in \cite{stevens95a}. All
beads and counterions interacted with the truncated Lennard-Jones
potential plus the full Coulomb interaction.

In this work experimental values of the osmotic pressure and the
maximum position in the interchain structure facture were successfully
reproduced. One of the important findings was that the chains
essentially are never rodlike\index{Rodlike}. Counterion-chain
correlations can dramatically shrink the polyelectrolyte chain.  The
end-to-end distance shortens significantly as the density increases
from dilute towards the overlap density.  The chain structure is
highly anisotropic in the very dilute limit, and the scaling with
respect to $N_{m}$ is asymmetric; but as the overlap density is
approached, the structural anisotropy dissipates and the scaling
becomes approximately symmetric. On long length scales the chain
structure continuously changes from very elongated to neutral-like
coils. Yet, on short length scales, the chain structure is density
independent and elongated more than neutral chains.

It was found that in the dilute limit the scaling for the extension
perpendicular to the chain was $R_{\bot} \propto N^{0.65 - 0.70}$, and for the
extension parallel $R_{\parallel} \propto N^{0.90 - 1.00}$. Near the density,
where the rodlike\index{Rodlike} chains in disordered solution  overlap,
$\rho \sim N^{-2}, R_{\bot}$ grows at the expense of $ R_{\parallel}$
until at the overlap density $\rho^{\star}$ the effective exponent is about
0.82. The transition regime ranges from $\rho \sim N^{-2}$ to about $\rho \sim
N ^{-1.4}$ where the coils start to overlap and one eventually reaches $\nu =
1/2$ in the semidilute regime. The exponents reported should not necessarily
be taken as asymptotic ($N \rightarrow \infty$), however they should be
relevant for many experimental systems.

\subsection{Poor solvent chains with explicit counterions}\label{subsec:poorsolvent}

\begin{figure}[tb]
  \begin{center}
    \leavevmode
    \epsfig{file=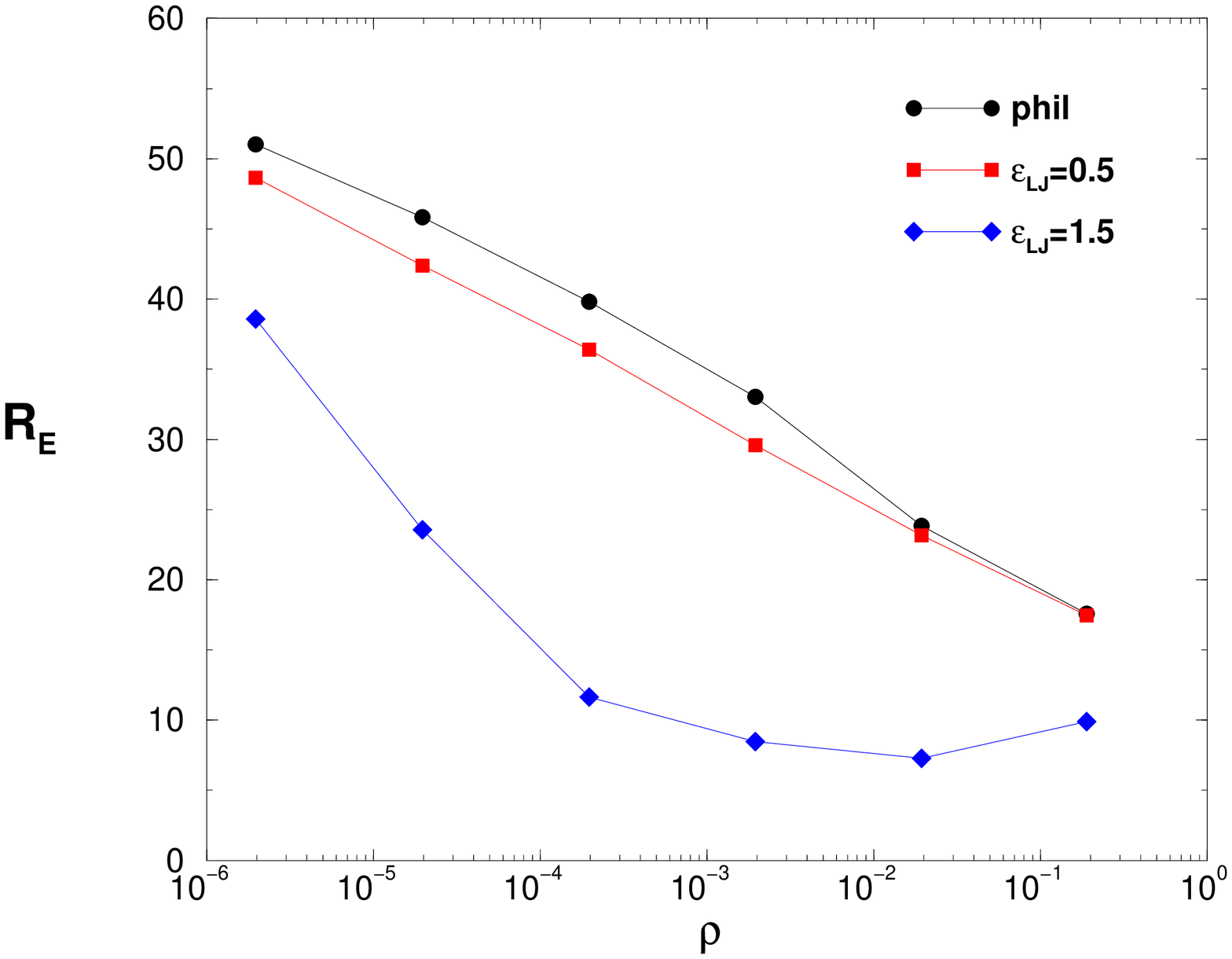,width=0.49\textwidth} 
    \epsfig{file=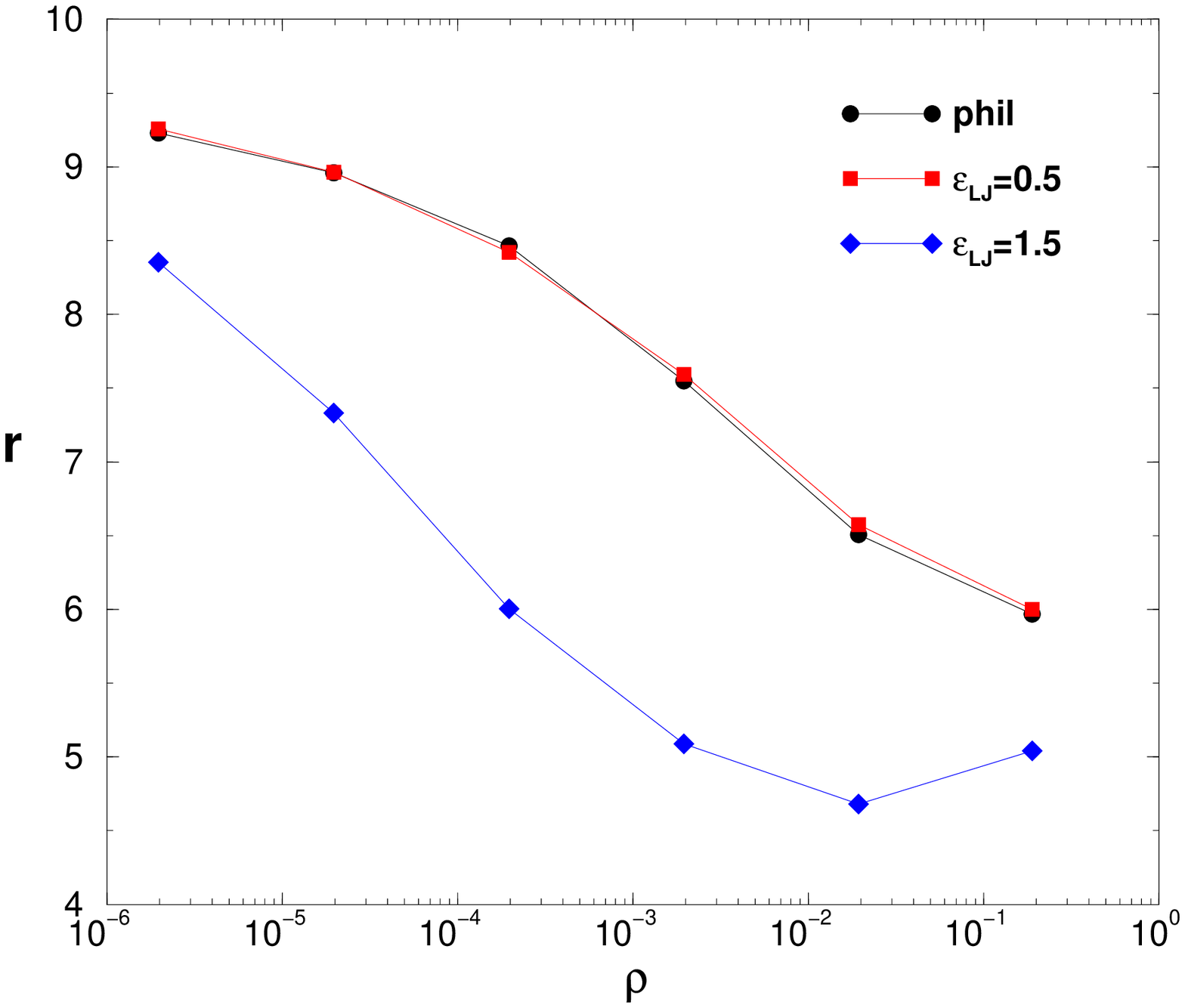,width=0.49\textwidth}
    \caption{ ${\rm R_E}$ (left) and r (right) versus density $\rho$ for
    hydrophilic 
    (phil), weak hydrophobic ($\epsilon_{LJ} = 0.5$), and strongly hydrophobic
    ($\epsilon_{LJ} =1.5$) chains} 
    \label{micka_fig_re}
  \end{center}
\end{figure}

\begin{figure}[tbp]
  \begin{center}
    \leavevmode
     \epsfig{file=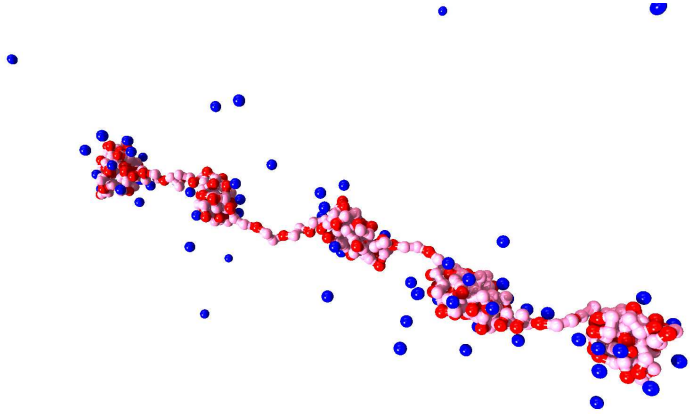,angle=90,width=\textwidth}
\caption{Typical polyelectrolyte conformation for a density $\rho =
     2\cdot10^{-4} \sigma^{-3}$, showing 5 pearls. The chain had 382 monomers
  with a charge fraction $f=1/3$, $l_B=1.5$, and $\epsilon = 1.75$.} 
    \label{fig:5pearl}
  \end{center}
\end{figure}

Many polyelectrolytes possess a carbon based backbone for which water is a
poor solvent. Therefore, in aqueous solution, there is a competition between
the solvent quality, the Coulombic repulsion, and the entropic degrees of
freedom. The conformation in these systems can under certain conditions
assume pearl-necklace like structures\cite{dobrynin96a}. These also exist for
strongly charged polyelectrolytes at finite densities in the presence of
counterions\cite{Holm:99,limbach01c}.  The simulations in Ref.\cite{Holm:99}
used 16 chains of length $ N_{m}$ = 94, with a charge fraction of $f$ = 1/3,
and monovalent counterions.  The hydrophobic interaction strength was tuned by
means of the Lennard-Jones parameter $\epsilon$. There we showed that the
polymer density $\rho$ can be used as a very simple parameter to separate
different conformation regimes. This can already be seen in the plots of the
end-to-end distance $R_e$ and $r= \frac{R_E^2}{R_G^2}$ versus $\rho$ in
Fig.~\ref{micka_fig_re}. At very high densities the electrostatic interaction
is highly screened, so that the hydrophobic interaction wins, and the chains
collapse to dense globules. If one slightly decreases the density, the chains
can even contract further, because there are no more steric hinderences from
the other chains or counterions, and the screening is smaller.  The collapsed
globules, however, have still a net charge, and repel each other, so that this
phase resembles a charged stabilized colloid or microgel phase. 
With
decreasing density the electrostatic interaction will dominate over the
hydrophobic one.  The chains will tend to elongate, assuming pearl-necklace
conformations, like in Fig.~\ref{fig:5pearl}, as they have been predicted for
weakly charged polyelectrolytes in Ref. \cite{dobrynin96a}.  The more the
chain stretches, the smaller become the locally compact regions.  Note that in
contrast to the analytical theories\cite{schiessel98a,dobrynin99a}, the pearls
are stable, even though there are counterions localized near and/or inside the
pearls.

%

\begin{figure}[tbp]\label{fig:pearlfluctuations}
  \begin{center}
     \epsfig{file=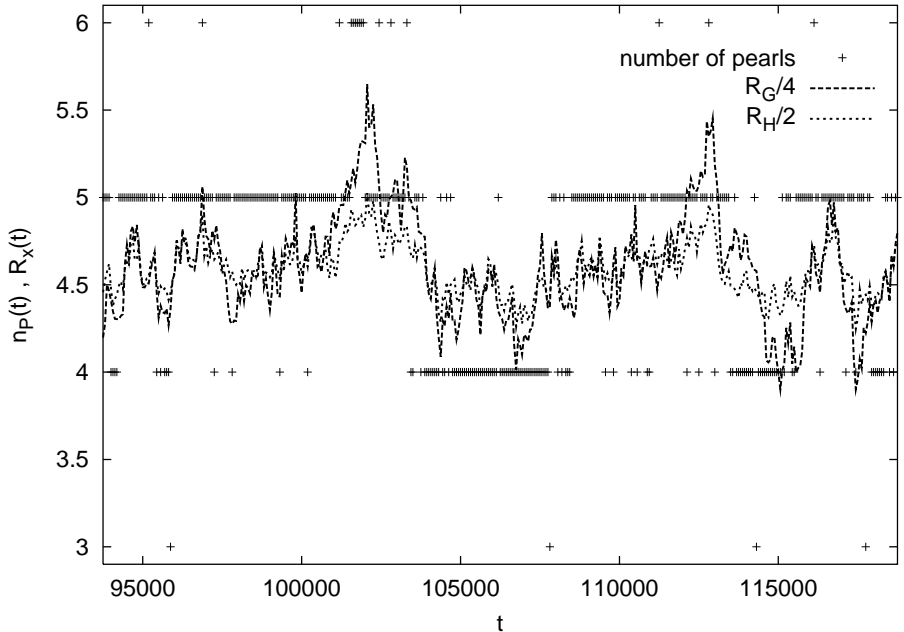,width=\textwidth}
\caption{Time development of the radius of gyration $R_G$, the hydrodynamic
  radius $R_H$ and the number of observed pearls for the same system as in
  Fig.~\ref{fig:5pearl}. }
  \end{center}
\end{figure}

Experimentally there are some hints for the existence of pearl-necklace
chains\cite{rawiso_lh,williams_lh}. One of the obstacles to observing them in
scattering experiments could be related to the strong fluctuations of the
pearl number. Even in equilibrium we have found coexistence of several pearl
states\cite{limbach01c}.  In Fig.~\ref{fig:pearlfluctuations} we see the time
evolution of one single chain composed of 382 monomers with a charge fraction
$f=1/3$, $l_B=1.5$, and $\epsilon = 1.75$ in a many chain system at density
$\rho = 1.48 \times 10^{-5}$. One observes jumps between a five and four pearl
configuration. Also the position of the pearls move quite vividly.

The different length scales appearing in a chain can be analyzed by looking at
the spherically averaged form factor $S_1(q)$ of the chain. 
\begin{figure}[tbp]
  \begin{center}
     \epsfig{file=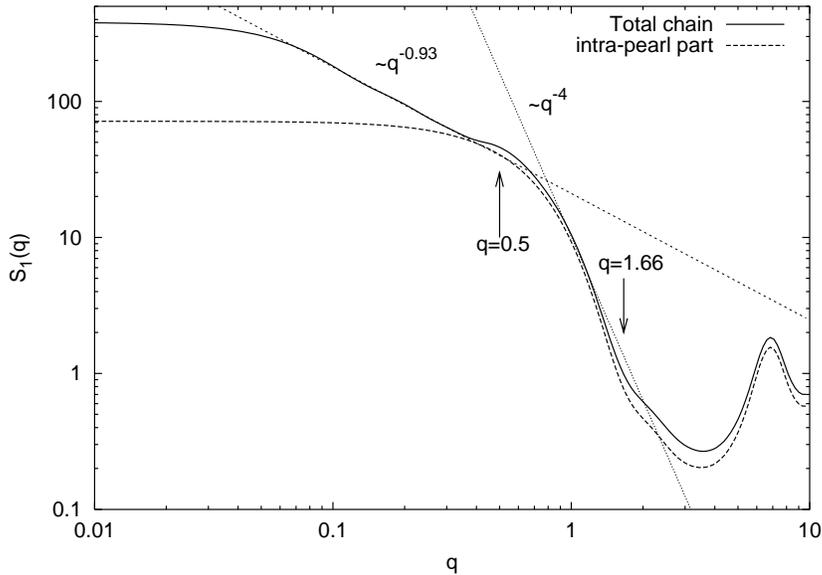,width=0.9\textwidth}
\caption{Spherically averaged form-factor $S_1(q)$ for the same system as in
  Fig.~\ref{fig:pearlfluctuations}The solid line denotes the single chain
  form-factor, the dashed line shows only the contributions of monomers within
  the same pearl.}
  \end{center}\label{fig:formfactor}
\end{figure}
The maximum seen at $q=6$ comes from the monomer extension. In the
range $1<q<2$ we observe a sharp decrease in $S_1$, which comes from
the scattering from the pearls, because it shows the typical Porod
scattering of $S_1(q) \simeq q^{-4}$. The kink at $q\approx 1.66$
appears at the position expected from the pearl size, but is broadly
smeared out due to large size fluctuations. The shoulder which can be
seen at $q\approx 0.5$ does not come from the intra-pearl scattering
but is due to the scattering of neighboring pearls along the chain
(inter-pearl contribution), which have a mean distance of $\langle
r_{PP}\rangle = 13.3.\sigma$. It is also smeared out due to large
distribution of inter-pearl distances. We conclude that the signatures
of the pearl-necklaces are weak already for monodisperse samples. A
possible improvement could be achieved for chains of very large
molecular weights and only few pearl numbers, which could lead to
stable and large signatures. Many more interesting results on poor
solvent polyelectrolytes can be found in Ref.\cite{limbach01c} and
will be published soon.

\subsection{Counterion distribution around finite polyelectrolytes}\label{sec:endeffects}
We recently completed a study of the spatial distribution of the
counterions around strongly charged, flexible
polyelectrolytes\cite{limbach01a} in good and poor solvent. There we
demonstrated that by partially neutralizing the quenched charged
distribution on the chain backbone the inhomogeneous distribution of
counterions leads to the same qualitative effects that are observed in
weakly charged polyelectrolytes with an annealed charge
distribution\cite{joanny_lh}.  This is due to the presence of the
mobile partially neutralizing counterions, which results in an
annealed backbone charge distribution. The common underlying physical
mechanism for the end-effect is the differences in the electrostatic
field of the chain along its backbone. The strength of the end-effect
depends on parameters like chain length, charge fraction and ionic
strength, and those dependencies were found in agreement with the
scaling predictions.  We found a saturation of the end-effect for long
chains, when the chain extension, namely $R_e$, is at least twice as
large as the Debye screening length.  A simple Debye-length criterion
appeared to be sufficient to explain the penetration depth of the
end-effect.  However, looking at the amplitude dependency on density
and ionic strength of the solution, we found that both parameters, the
number of annealing ions and the ionic strength of the solution,
influence the end-effect and that the first one dominated.  The
amplitude of the end-effect was shown to depend strongly on the charge
parameter $\xi:=l_B/b$, where $b$ is the distance of the bare charges
on the backbone of the chain. The definition of such an end-effect via
close mobile counterions can not be made for an effective charge $\xi
<< 1$, because under dilute conditions there are almost no counterions
close to the chain.

Even though the chain conformation is very different in the poor
solvent case the end-effect was found to be qualitatively the same,
namely the counterions are more likely to be found at the middle of
the chain than at the ends.  We could also clearly see the necklace
structure by looking at the effective charge along the contour
length. However, the string length of our simulated pearl-necklaces
was too short to show any charge difference between the pearls and the
strings, as has been predicted in Ref. \cite{castelnovo00a}.

We also obtained a fairly good agreement of the simulated ion
distribution with the PB solution of the cell model of an infinitely
extended charged rod\cite{deserno_lh}. This supports the idea that the
description of polyelectrolytes as rodlike objects in mean-field
theory is valid in the dilute regime. Further improvements could
probably be achieved along the lines of Ref.\cite{deshkovski01a},
where a combination of a cylindrical and spherical cell model is used
to describe the solution properties of polyelectrolytes.

\subsection{Rodlike\index{Rodlike} polyelectrolytes}\label{sec:rods}

Stiff linear polyelectrolytes can be approximated by charged
cylinders. This is a relevant special case, applying to quite a few
biologically important polyelectrolytes with a large persistence
length, like DNA, actin filaments or microtubules.  Within PB theory
\cite{gouy10a} and on the level of a cell model the cylindrical
geometry can be treated exactly in the salt-free case
\cite{deserno_lh,alfrey51a,katchalsky71a,joensson_lh,lebret84a,lebret84b,deserno00a},
providing for instance new insights into the phenomenon of the Manning
condensation \cite{manning69a,oosawa71a}. For low line charges, the
agreement between PB theory and the simulations of the full
interaction system is rather nice. However, PB theory fails
quantitatively (underestimated condensation) and qualitatively
(overcharging\index{Overcharging}, charge oscillations and attractive
interactions); see, e.g.  Ref.\cite{deserno00a,deserno00b,deserno00c}.

Recently the osmotic coefficient of a synthetic stiff polyelectrolyte,
a poly(para-phenylene), was measured in a salt-free
environment\cite{guilleaume00a,blaul00a}. We have compared this data
to predictions of PB theory, and a local density functional theory
which includes a correlation correction of the basis of a recently
proposed Debye-H\"uckel-Hole-cavity theory (DHHC) \cite{barbosa00a},
and simulational results within the cell model. We find that
correlation effects enhance condensation and lower the osmotic
pressure, yet are not fully able to explain the discrepancy with the
experimental data. Here the approach of working within the ``primitive
model'' breaks down. In our opinion, specific interactions between the
counterions, the macroion, and the solvent particles are needed to
explain the discrepancy. Other theoretical approaches beyond the cell
model which try to incorporate finite-size effects and interactions of
the macroion itself will in general lead to a higher osmotic
coefficient which is in contrast to the experimental
data\cite{deserno01a}.

Attractive interactions have been
observed\cite{podgornik94a,tang96a,bloomfield96a,lyubartsev98a} and
predicted between like-charged macromolecules. However, there are nice
rigorous results which prove that these effects cannot be described by
mean-field theories\cite{neu99a,sader99a,trizac00a}.  Especially in
the community of biological inspired
physics\cite{podgornik_lh,gelbart_lh,khokhlov_lh,nguyen_lh}, these
interactions are thought to be important for the clarification of the
mechanism behind DNA compactification in viral heads\cite{lambert00a},
the chromatin structure\cite{holde89a}, and novel methods for gene
delivery\cite{kabanov95a}, to name just the most prominent
examples. There are numerous simulations which show similar
attractions\index{Attractions} on a distance of few counterion
diameters
\cite{deserno00b,guldbrand84a,nilsson91a,lyubartsev97a,jensen97a,stevens99a,allahyarov00a}.

The mechanism which is driving the observed attractions for rod-like systems
has been speculated to be correlations between the counterion layers around
the macroion. However, until now, no unique theoretical picture has emerged
that can clarify the detailed mechanism behind the attractions. There is the
low temperature Wigner crystal theory, initiated by Refs.
\cite{rouzina96a,shklovskii99a,nguyen_lh}, which postulates an ordered ground
state of the counterions. Then there are theories which are based on Van der
Waals type correlated fluctuations
\cite{joensson_lh,oosawa68a,spalla95a,ha97a}, that are in principle hight $T$
theories. There are also theories which are fluctuation based, but are valid
at low $T$\cite{moreira_lh,lau00a,lau00b}. Integral equation
\cite{kjellander_lh,gonzalestovar85a,kjellander84a} theories on various
approximation levels have been demonstrating the existence of these
attractions for a long time, but from these theories it is difficult to
extract the detailed mechanism behind the observed correlations.  Here also
simulations can be helpful, because they have in principal access to all
correlations\cite{deserno01d}.  More details of our results in rod-like
geometries can be found in
Refs.\cite{deserno00a,deserno00b,deserno00c,barbosa00a,deserno01a,deserno00f,deserno01b}.

\section{The energetic path to understand overcharging\index{Overcharging}}\label{sec:overcharge}
There has been a recent interest in the study of systems which are strongly
coupled by Coulomb interactions. These systems show a variety of, at first
sight, surprising behaviors, which can not be accounted for by the mean-field
PB theory. For example, there are attractions\index{Attractions} between like charged objects and
a charge reversal of macroions occurs when viewed from some distance. This means
that there are more ions of the opposite charge within a certain radius around
the macroion then necessary to charge neutralize it. This
overcompensation is called ``overcharging''.

In this section we want to demonstrate that there are situations for charged
colloidal objects in which one can understand the phenomenon of overcharging
by very simple energetic arguments. By overcharging, in general, we mean that
the bare charge of the macroion is overcompensated at some distance by
oppositely charged ``microions''. To achieve this in nature we have to add salt to
the system. For the sake of simplicity, however, we will consider non-neutral
systems, because they can on a very simple basis explain why colloids prefer to
be overcharged. 

\subsection{The Model}
Our model is solely based on electrostatic energy considerations, meaning that
we only look at the ground state of a system of charges.  We consider a
colloid of radius $a$ with a central charge $Z$. In the ground state the
counterions of this colloid are located on the surface, because there they are
closest to the central charge. On the other hand they want to be in such a
configuration that they minimize their mutual repulsion. For two, three, and
four counterions these configurations correspond to a line, an equilateral
triangle, and a tetrahedron, respectively, regardless of the central
charge magnitude.  The problem of the minimal energy configuration of
electrons disposed on the surface of a sphere dates back to Thomson
\cite{thomson04a}, and is actually unsolved for large $N$. The
reason is, that there are many metastable states which differ only minimally in
energy, and their number seems to grow exponentially with $N$. Also chemists
developed the valence-shell electron-pair repulsion (VSEPR) theory
\cite{oxtoby99a} which uses similar arguments to predict the molecular
geometry in covalent compounds, also known as the Gillespie rule.

\begin{figure}
\epsfig{file=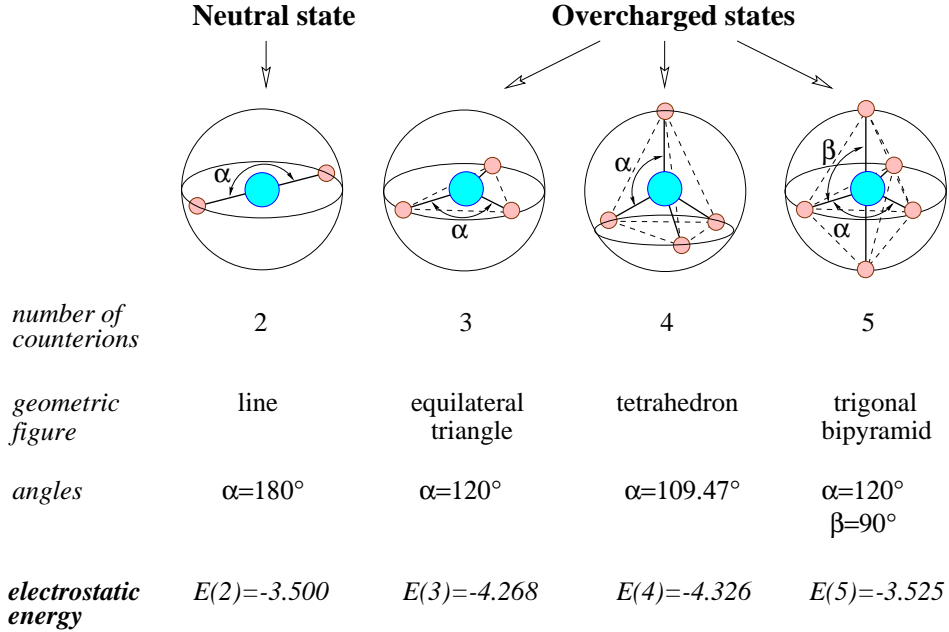,width=\textwidth}
\caption{Ground state configurations for two, three, four and five
  electrons. The corresponding 
geometrical figure repulsion and their typical angles are given. The electrostatic
energy (in units of \protect\( k_{B}Tl_{B}/a\protect \)) is given for a central
charge of +2\textit{e}. }
\label{fig:gillespie} 
\end{figure}

A simple illustration of energetically driven overcharging is depicted in
Fig.~\ref{fig:gillespie}. The central charge is +2\textit{e}, and the neutral
system has two counterions of valence 1. If we add successively more
counterions of the same valence, and put them on the surface such that their
mutual repulsion is minimized, we can compute the total electrostatic energy
according to
\begin{eqnarray}
\label{Eq.gillespie}
E(n)=k_{B}T(l_{B}/a)\left[ -nZ_{m}+f(\theta_i )\right] ,
\end{eqnarray}
where \( f(\theta _i) \) is the repulsive
energy part which is only a function of the 
ground state configuration. We surprisingly find, that actually the minimal
energy is obtained when {\it four} counterions are present, hence we
overcharged the colloid by two counterions, or by 100 \%! That is, the excess
counterions gain more energy by assuming a energetically favorable configuration
around the macroion than by escaping to infinity, the simple reason behind
overcharging.
In our example, the minimum is reached when four counterions are present. The
colloid radius and the Bjerrum length enter as prefactors and change only the
energy difference between neighboring states.

The spatial correlations of the counterions are fundamental to obtain
overcharging. Indeed, if we apply the same procedure and smear \( Z \)
counterions onto the surface of the colloid of radius \( a \), we obtain for
the energy
\begin{eqnarray}
\label{smeared}
E=l_{B}\left[ \frac{1}{2}\frac{Z^{2}}{a}-\frac{Z_{m}Z}{a}\right].
\end{eqnarray}
The minimum is reached for \( Z=Z_{m} \), hence no overcharging can occur.
  
The important message to be learned is that, from an energetic point of
view, a colloid \textit{always} tends to be overcharged by discrete charges.
Other important geometries like infinite rods or infinitely extended plates
cannot be treated in such a simple fashion because they are not finite in all
directions. One needs therefore enough screening charges in the environment to
limit the range of the interactions in the infinite directions, there is a
need for a {\it minimal} amount of salt present to allow for
overcharging\cite{deserno01b}, which is not the case for a colloid.
 
Obviously, for a large number of counterions the direct computation of the
electrostatic energy by using the exact equation (\ref{Eq.gillespie}) becomes
unfeasible.  Therefore we resort to simulations for highly charged
spheres.

\subsection{One colloid}
\begin{figure}
  \centerline{\epsfig{file=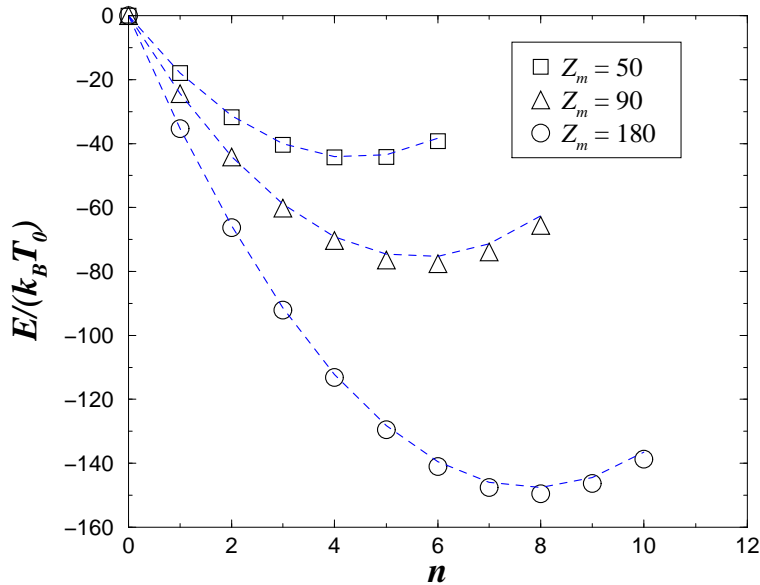,width=0.8\textwidth}}
\caption{Electrostatic energy (in units of \protect\( k_{B}T_{0}\protect \))
  for \textit{ground state} configurations of a single charged macroion as
  a function of the number of \textit{overcharging} counterions \protect\(
  n\protect \) for three different bare charges \textit{\protect\(
    Z_{m}\protect \)}. The neutral case was chosen as the potential energy
  origin, and the curves were produced using the theory of Eq.
\  (\ref{Eq.WC-n-OC}), compare text.}
\label{fig.OC-MD-energy}
\end{figure}

The electrostatic energy as a function of the number of overcharging
counterions \( n \) is displayed in Fig.~\ref{fig.OC-MD-energy}. We note that the
maximal (critical) acceptance of \( n \) (4, 6 and 8) increases with the
macroionic charge \( Z_{m} \) (50, 90 and 180 respectively). Furthermore for
fixed \( n \), the gain in energy is always increasing with \( Z_{m} \).
Also, for a given macroionic charge, the gain in energy between two successive
overcharged states is decreasing with \( n \).

In the ground state the counterions are highly ordered. Rouzina and Bloomfield
\cite{rouzina96a} first stressed the special importance of these crystalline
arrays for interactions of multivalent ions with DNA strands, and later
Shklovskii \cite{nguyen_lh} showed that the Wigner crystal (WC) theory can be
applied to determine the interactions in strongly correlated systems. In two
recent short contributions \cite{messina00a,messina00b} we showed that the
overcharging curves obtained by simulations of the ground state, like
Fig.~\ref{fig.OC-MD-energy}, can be simply explained by assuming that the
energy \( \varepsilon \) per counterion on the surface of a macroion scales as
$\sqrt{c}$, where $c$ denotes the counterion concentration $c=N/A$, $N$ is the
\textit{total} number of counterions on the surface and $A$ the total macroion
area.  This can be justified by a simple argument, where each ion interacts in
first approximation only with the oppositely charged background of its
Wigner-Seitz (WS) cell, which can be approximated by a disk of radius \( h \),
yielding the same WS cell area.

For fixed macroion area we can write the energy per counterion as
\begin{eqnarray}
\label{eq.E-N}
\varepsilon ^{(h)}(N)=-\frac{\bar \alpha ^{(h)}\ell }{\sqrt{A}}\sqrt{N} = 
- \bar \alpha ^{(h)} \ell \sqrt{c},
\end{eqnarray}
where \( \ell =l_{B}Z_{c}^{2} \) and the simple hole theory gives \(
\bar \alpha ^{(h)}=2\sqrt{\pi }\approx 3.54 \)\cite{messina01a}.

For an infinite plane, where the counterions form an exact triangular lattice,
one obtains the same {\it functional} form as in Eq. (\ref{eq.E-N}), but the
prefactor \( \bar \alpha ^{(h)} \) gets replaced by the numerical value \(
\bar \alpha
^{WC}=1.96 \)\cite{bonsall77a}.

Not knowing the precise value of $\bar \alpha$ we can still use the simple
scaling behavior with \( c \) to set up an equation to quantify the energy
gain \( \Delta E_{1} \) by adding the first overcharging counterion to the
colloid. To keep the OCP neutral we imagine adding a homogeneous surface
charge density of opposite charge ($\frac{-Z_c e}{A}$) to the
colloid\cite{nguyen_lh}. This 
ensures that the background still neutralizes the incoming overcharging
counterion and we can apply Eq. (\ref{eq.E-N}). To cancel our surface charge
addition we add another homogeneous surface charge density of opposite sign
$\frac{Z_c e}{A}$. This surface charge does not interact with the now neutral
OCP, but adds a self-energy term of magnitude $\frac{1}{2}\frac{\ell}{a}$, so
that the total energy difference for the first overcharging counterion reads as
 \begin{eqnarray}
\label{Eq.WC-FIRST-OC}
\Delta E_{1}=(N_{c}+1)\varepsilon (N_{c}+1)-N_{c}\varepsilon (N_{c}) + \frac{\ell}{2a}.
\end{eqnarray}
By using Eq. (\ref{eq.E-N}) this can be rewritten as\cite{messina01b}

\begin{eqnarray}
\label{Eq.WC-FIST-OC-b}
\Delta E_{1}=-\frac{\bar \alpha \ell }{\sqrt{A}}\left[
  (N_{c}+1)^{3/2}-N_{c}^{3/2}\right] + \frac{\ell}{2a}.
\end{eqnarray}
Completely analogously one derives for
the energy gain \( \Delta E_{n} \) for \( n \) overcharging counterions
\begin{eqnarray}
\label{Eq.WC-n-OC}
\Delta E_{n}=-\frac{\bar \alpha \ell }{\sqrt{A}}\left[ (N_{c}+n)^{3/2}-N_{c}^{3/2}\right] +\frac{\ell }{a}\frac{n^2}{2}.
\end{eqnarray}
%
%
Using Eq.  (\ref{Eq.WC-n-OC}), where we determined the unknown \( \bar \alpha
\) from the simulation data for \( \Delta E_{1} \), we obtain a curve that
matches the simulation data almost perfectly (Fig.~\ref{fig.OC-MD-energy}).
The second term in Equation (\ref{Eq.WC-n-OC}) also shows why the overcharging
curves of Fig.~\ref{fig.OC-MD-energy} are shaped parabolically upwards for
larger values of \( n \).  If one successively removes each of $n$ counterions
from a neutral colloid, one can derive in a similar fashion the ionization
energy cost
\begin{eqnarray}
\label{Eq.ionization}
\Delta E_{n}^{ion}=-\frac{\bar \alpha \ell }{\sqrt{A}}\left[ (N_{c}-n)^{3/2}-N_{c}^{3/2}\right] +\frac{\ell }{a}\frac{n^2}{2}.
\end{eqnarray}
Using the measured value of \( \bar \alpha \) we can simply determine the
maximally obtainable number \( n_{max} \) of overcharging counterions by
finding the stationary point of Eq. (\ref{Eq.WC-n-OC}) with respect to \( n
\):
\begin{eqnarray}
\label{Eq.Q*}
n_{max}=\frac{9 \bar \alpha ^{2}}{32\pi }+
\frac{3\bar \alpha }{4\sqrt{\pi }}\sqrt{N_{c}}
\left[ 1+ \frac{9 \bar \alpha^2}{64\pi N_c }\right] ^{1/2}.
\end{eqnarray}
The value of \( n_{max} \) depends only on the number of counterions
\( N_{c} \) and \( \bar \alpha \). For large \( N_{c} \)
Eq. (\ref{Eq.Q*}) reduces to \( n_{max}\approx \frac{3 \bar \alpha
}{4\sqrt{\pi }}\sqrt{N_{c}} \) which was derived in
Ref. \cite{shklovskii99b} as the low temperature limit of a a neutral
system in the presence of salt. What we have shown is that the
overcharging in this limit has a pure electrostatic origin, namely it
originates from the energetically favorable arrangement of the ions
around a central charge.  We also showed in Ref. \cite{messina01b}
that $ \bar \alpha$ reaches the perfect WC value of $1.96$ if the
colloid radius $a$ gets very large at fixed $c$, or when $c$ becomes
large at fixed $a$.

If instead of a central charge scheme one uses discrete charge centers
distributed randomly over the colloidal surface we find counterion
structures which are quite far away from the WC array, especially when
the counterions are pinned to their counter charges. This depends on
the interaction energy at contact, which depends of course on $l$ and
distance of closest approach.  However, we still find overcharging,
although reduced in value, of the form given by Eq. \ref{Eq.WC-n-OC}
\cite{messina01a,messina01c}

\subsubsection{Macroion-counterion interaction profile at $T = 0 K$ \label{sec.Interaction-profile}}

The interaction profile between a completely neutralized macroion and
one excess counterion is obtained by displacing adiabatically the
excess counterion from infinity towards the macroion. From far away
the counterion sees only a neutral object and has no measurable
interaction, whereas upon approach to the macroion the WC hole gets
created in the counterion layer, and we observe a distance dependant
attraction towards the macroion. We investigated cases of \( Z_{m}=2
\dots 288 \). All curves can be nicely fitted with an exponential fit
of the form
\begin{eqnarray}
\label{eq.Interaction-fit}
E_{1}(r)=\Delta E_{1}e^{-\tau (r-a)},
\end{eqnarray}
where \( \Delta E_{1} \) is the measured value for the first
overcharging counterion, and \( \tau \) is the only fit parameter. In
all our results for \( \tau \) versus \( \sqrt{N_{c}} \) we observe a
linear dependence for a wide range of values for \( N_{c} \), $\tau
\propto \sqrt{N_{c}}$, which again can be explained by applying the WC
hole picture \cite{messina01b}.

\subsection{Two Colloids}
Now we apply what we have learned about a single colloid to two equal-sized,
fixed charged spheres of bare charge $Q_{A}$ and $Q_{B}$ separated by a
center-center separation $R$ and surrounded by their neutralizing counterions,
which give concentrations $c_A$ and $c_B$, respectively.

All these ions
 making up the system are immersed in a cubic box of length \(
L=80\sigma \), and the two macroions are held fixed and disposed
symmetrically along the axis passing through the centers of opposite
faces. This leads to a colloid volume fraction \( f_{m}=2\cdot
\frac{4}{3}\pi (a/L)^{3}\approx 8.4\times 10^{-3} \). For
\textit{finite} colloidal volume fraction \( f_{m} \) and temperature,
we know from the study carried out above that in the strong Coulomb
coupling regime all counterions are located in a spherical
{}``monolayer{}'' in contact with the macroion.  Here, we investigate
the mechanism of \textit{strong, long range} attraction stemming from
\textit{monopole} contributions; that is, one colloid is overcharged
and the other one undercharged.

\subsubsection{Observation of metastable ionized states\label{sec.obsevation-metastable-states}}

For the charge symmetrical situation we have $c_{A}=c_{B}$. When we brought
this system to room temperature $T_{0}$ and generated initially the
counterions randomly inside the box we observed in some cases that one of the
colloids remained undercharged, and the other one was overcharged, and these
configurations turned out to be extremely long lived in the course of our MD
simulations( more than \( 10^{8} \) MD time steps).  However it is clear that
such a state is ``metastable'' because by symmetry arguments it cannot be the
lowest energy state.  The observed barrier is the result of the WC attraction,
because close to the macroion surface the energy is reduced. For very distant
macroions the barrier height for the first overcharged state has to equal
$\Delta E_1$ from Eq. \ref{Eq.WC-n-OC}. The barrier profile at $T=0$ can also
be extremely well approximated by an application of Eqs. (\ref{Eq.WC-n-OC})
and (\ref{Eq.ionization}), plus taking into account the distant dependent
monopole contribution\cite{messina00a}. This leads to a barrier height which
scales as $\sqrt{c}$ for large separations. For smaller separations one has to
take into account also the effect of strong mutual polarization of both
macroions, which leads effectively to a sharing of their proximal counterion
layer into a superlattice. This can be taken into account by a higher
effective counterion density close to the surface, leading to an almost linear
scaling of the barrier height with $c$ \cite{messina00a,messina01b}.

\subsubsection{Asymmetrically charged colloids\index{Charged colloid}}

The most interesting phenomenon, however, appears when the two colloids have
different counterion concentrations, here $c_A > c_B$, since then {\bf stable ionized states} can
appear. The physical reason is that a counterion can gain more energy by
overcharging the colloid with $c_A$ then it loses by ionizing colloid $B$. A
straight forward application of the procedure outlined for the barrier
calculation \cite{messina00b,messina01b} yields a
simple criterion (more specifically a sufficient condition), valid for large
macroionic separations, for the charge asymmetry \( \sqrt{N_{A}}-\sqrt{N_{B}}
\) to produce an ionized ground state of two unlike charged colloids with the
same size:
\begin{eqnarray}
\label{Eq.criterion}
\left( \sqrt{N_{A}}-\sqrt{N_{B}}\right) >\frac{4\sqrt{\pi }}{3\bar \alpha ^{A}}
\approx 1.2. 
\end{eqnarray}

\subsubsection{Finite temperature analysis}

\begin{figure}
\epsfig{file=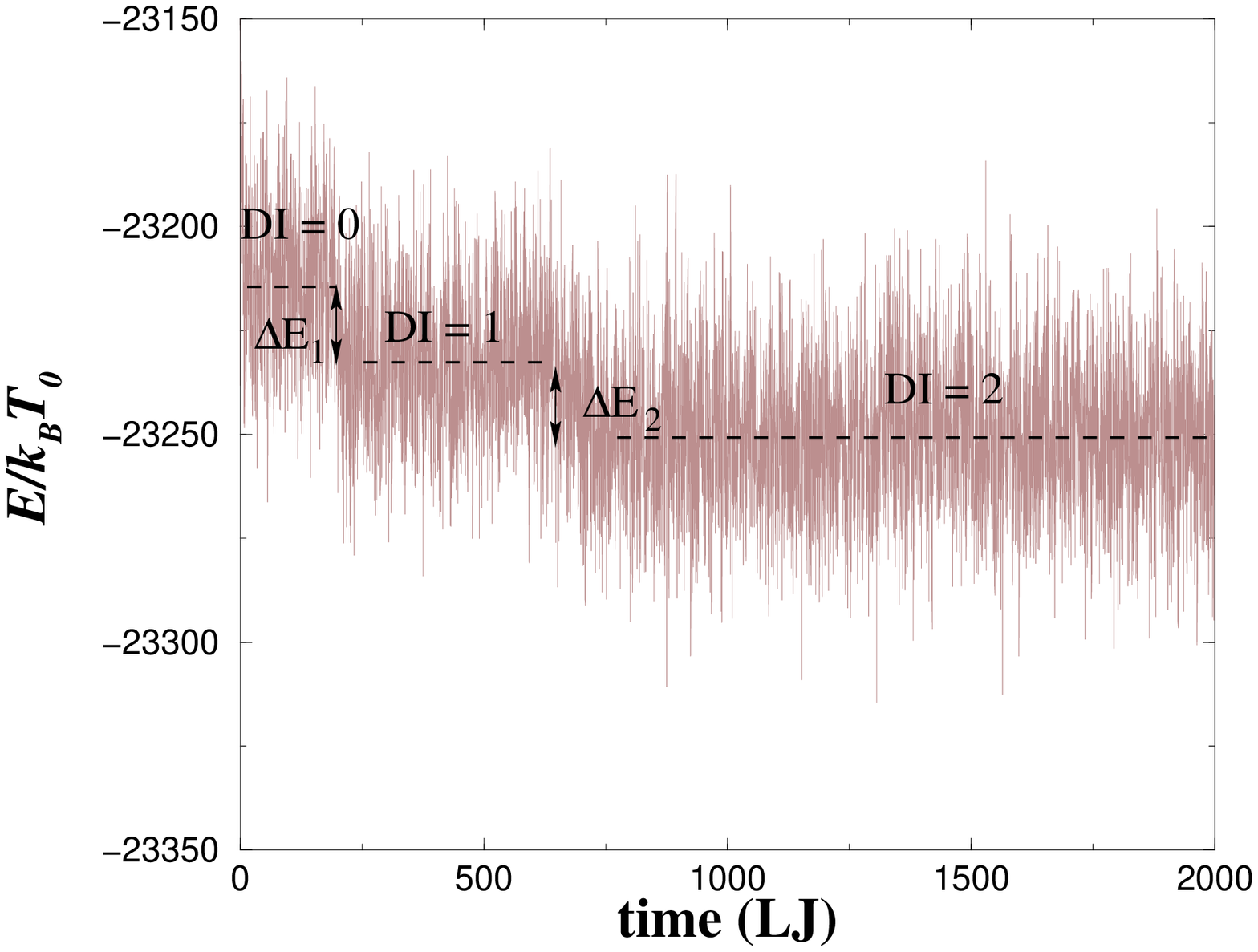,width=0.49\textwidth}
\epsfig{file=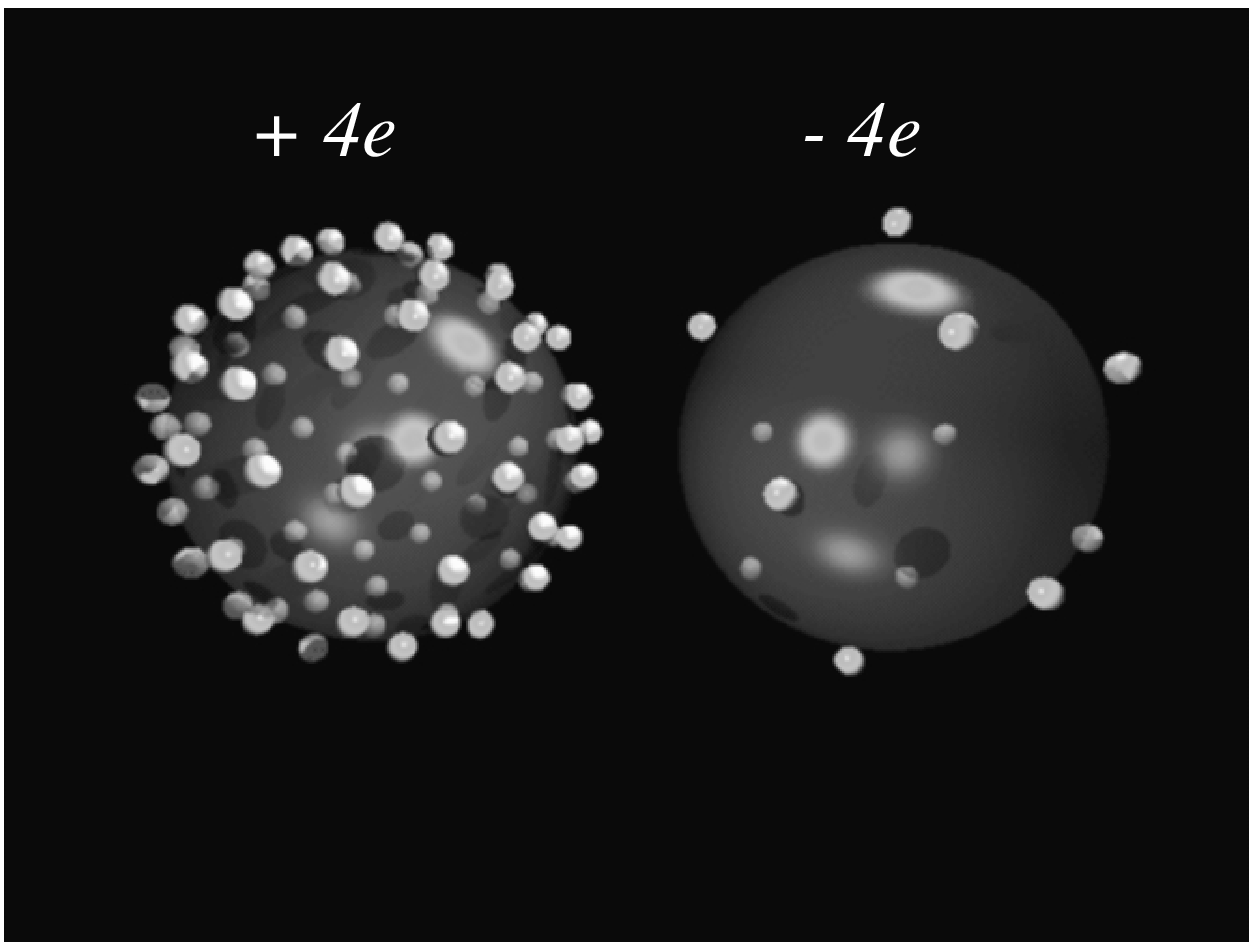,width=0.49\textwidth}
\caption{Relaxation, at room temperature \protect\( T_{0}=298K\protect \), of
  an initial unstable neutral state towards ionized state. Plotted is the
  total electrostatic energy versus time (LJ units), for \protect\(
  Z_{B}=30\protect \) and \protect\( R/a=2.4\protect \).  Dashed lines lines
  represent the mean energy for each \textit{DI} state. Each jump in energy
  corresponds to a counterion transfer from the macroion \textit{B} to
  macroion \textit{A} leading to an ionized state that is lower in energy than
  the neutral one. The right figure is a snapshot of the final ionized state,
  with net charges {\it +4e} and {\it -4e} as indicated.}
\label{fig.relaxation}
\end{figure}

We have also demonstrated that the ground state phenomena survive for finite
temperatures, i.e. an ionized state can also exist at room temperature \(
T_{0} \).  The left part of Figure \ref{fig.relaxation} shows the time
evolution of the electrostatic energy of a system \( Z_{A}=180 \) with \(
Z_{B}=30 \), \( R/a=2.4 \) and a colloidal volume fraction of $7\cdot
10^{-3}$, where the starting configuration is the neutral state (\textit{DI} =
0). One clearly observes two jumps in energy, \( \Delta E_{1}=-19.5\, \) and
\( \Delta E_{2}=-17.4\, \), which corresponds each to a counterion transfer
from colloid \textit{B} to colloid \textit{A}.  These values are consistent
with the ones obtained for the ground state, which are\( -20.1\, \) and \(
-16.3\, \) respectively. Note that this ionized state (\textit{DI} = 2) is
more stable than the neutral but is expected to be metastable, since it was
shown previously that the most stable ground state corresponds to \textit{DI}
= 5. The other stable ionized states for higher \textit{DI} are not accessible
with reasonable computer time because of the high energy barrier made up of
the correlational term and the monopole term which increase with \textit{DI}.
In the right part of Fig.~\ref{fig.relaxation} we display a typical snapshot
of the ionized state (\textit{DI} = 2) of this system at room temperature.

Obviously, these results are not expected by the DLVO theory even in the
asymmetric case (see e. g. \cite{daguanno92a}). Previous simulations of
asymmetric (charge and size) spherical macroions \cite{allahyarov98b} were
also unable to predict such a phenomenon since the Coulomb coupling was weak
(water, monovalent counterions). Note that the appearance of
(meta-)stable ionized states can alter the effective interactions between
charged colloids in solution. The monopole attraction will lead to attraction
between like charged colloids, flocculation, and related phenomena.

At this stage, we would like to stress again, that the appearance of a stable
ionized ground state is due merely to correlation. An analogous consideration
with smeared out counterion distributions along the lines of Eq.
(\ref{smeared}) will again always lead to two colloids exactly neutralized by
their counterions \cite{schiessel00b}.  Our energetic arguments are quite
different from the situation encountered at finite temperatures, because in
this case even a PB description would lead to an asymmetric counterion
distribution. However, in the latter case this happens due to purely entropic
reasons, namely in the limit of high temperatures, the counterions want to be
evenly distributed in space, leading to an effective charge asymmetry. 

Note also, that there can exist parameter regions, such as high molar
electrolytes, where the overcharging of a single macroion is due to mainly
entropic effects \cite{gonzalestovar85a,greberg98a,lozada01a}, whose exact
mechanism is currently under investigation\cite{messina01d}.

\section{Acknowledgments}
We gratefully acknowledge collaborations at various stages with A. Arnold,
J. DeJoannis, M. Deserno, H.J. Limbach, R. Messina, U. Micka, and Z. Wang.


\end{document}